\documentclass[11pt]{article}

\usepackage{epsfig,color,amsmath,times}

\def\nofull{0}
\def\plusminus{0}
\def\conf{0}

\ifnum\nofull=0
\usepackage{fullpage}
\else
\usepackage{full}
\usepackage{algorithmic}
\fi

   \newtheorem{thm}{Theorem}[section]
    \newtheorem{lemma}[thm]{Lemma}
   \newtheorem{corollary}[thm]{Corollary} \newtheorem{claim}[thm]{Claim}
   
   \newtheorem{definition}{Definition} \newtheorem{remark}{Remark}
   \newtheorem{alg}{Algorithm}

\newcommand{\BT}{\begin{thm}} \newcommand{\ET}{\end{thm}}
\newcommand{\BL}{\begin{lemma}} \newcommand{\EL}{\end{lemma}}
\newcommand{\BCM}{\begin{claim}} \newcommand{\ECM}{\end{claim}}
\newcommand{\BD}{\begin{definition}} \newcommand{\ED}{\end{definition}}
\newcommand{\BA}{\begin{alg}} \newcommand{\EA}{\end{alg}}

\newcommand{\BE}{\begin{enumerate}} \newcommand{\EE}{\end{enumerate}}
\newcommand{\BI}{\begin{itemize}} \newcommand{\EI}{\end{itemize}}

\def\FullBox{\hbox{\vrule width 8pt height 8pt depth 0pt}}
\newcommand{\qed}{\;\;\;\FullBox}
\newenvironment{proof}{\noindent{\bf Proof:~~}}{\(\qed\)}

\newcommand{\BEQ}{\begin{equation}} \newcommand{\EEQ}{\end{equation}}
\newcommand{\BEQN}{\begin{eqnarray}}\newcommand{\EEQN}{\end{eqnarray}}
\newcommand{\BPF}{\begin{proof}} \newcommand {\EPF}{\end{proof}}
\newenvironment{proofof}[1]{\noindent{\bf Proof of {#1}:~~}}{\(\qed\)}
\newcommand{\BPFOF}{\begin{proofof}} \newcommand {\EPFOF}{\end{proofof}}

\newcommand{\eps}{\epsilon}
\newcommand{\poly}{{\rm poly}}
\newcommand{\polylog}{{\rm polylog}}
\newcommand{\eqdef}{\stackrel{\rm def}{=}}
\newcommand{\bitset}{\{0,1\}}
\renewcommand{\Pr}{{\rm Pr}}
\newcommand{\E}{{\rm E}}
\newcommand{\xth}{{\rm th}}

\begin{document}

\begin{titlepage}
\title{Approximating the influence of a monotone Boolean function in $O(\sqrt{n})$ query complexity}

\author{
Dana Ron\thanks{
School of Electrical Engineering at Tel Aviv University,
{\tt danar@eng.tau.ac.il}.
This work was supported by the Israel Science Foundation (grant number 246/08).}
\and
Ronitt Rubinfeld \thanks{
CSAIL at MIT, and the
Blavatnik School of Computer Science at Tel Aviv University,
{\tt  ronitt@csail.mit.edu}.
This work was supported by 
NSF grants 0732334 and 0728645,
Marie Curie Reintegration grant PIRG03-GA-2008-231077
and the Israel Science Foundation grant nos. 1147/09
and  1675/09}
\and
Muli Safra\thanks{
Blavatnik School of Computer Science at Tel Aviv University,
 {\tt safra@post.tau.ac.il}.}
\and
Omri Weinstein \thanks{
Computer Science Department,
Princeton University, {\tt oweinste@cs.princeton.edu}.}
}

\maketitle

\begin{abstract}

\sloppy
The {\em Total Influence\/} ({\em Average Sensitivity\/}) of
a discrete function is one of its fundamental measures. We study the
problem of approximating the total influence of a monotone Boolean function
\ifnum\plusminus=1
$f: \{\pm1\}^n \longrightarrow \{\pm1\}$,
\else
$f: \bitset^n \to \bitset$,
\fi
which we denote by $I[f]$.
We present a randomized algorithm that approximates the influence of such
functions to within a multiplicative factor of $(1\pm \eps)$
by performing $O\left(\frac{\sqrt{n}\log n}{I[f]}
\poly(1/\eps) \right) $ queries.
We also prove a lower bound of
$\Omega\left(\frac{\sqrt{n}}{\log n \cdot I[f]}\right)$
on the query complexity of any constant-factor approximation algorithm for
this problem (which holds for
$I[f] = \Omega(1)$),  
hence showing that our algorithm is almost optimal in terms of
its dependence on $n$.
For general functions we give a lower bound of
$\Omega\left(\frac{n}{I[f]}\right)$, which matches the complexity
of a simple sampling algorithm.

\end{abstract}
\noindent
\indent \hspace{.2cm} {\bf Keywords:}   influence of a Boolean function, sublinear approximation algorithms, random walks.

\end{titlepage}

\section{Introduction}


The influence of a function, first introduced by Ben-Or and Linial \cite{BL}
in the context of ``collective coin-flipping'', captures the notion of the
sensitivity of a multivariate function. More precisely, for a Boolean function
$f: \bitset^n \to \bitset$, the {\it individual influence\/}  of coordinate $i$ on $f$
is defined as $I_i[f] \eqdef \Pr_{x\in \bitset^n}[f(x) \neq f(x^{(\oplus i)})]$,
where $x$ is selected uniformly\footnote{The influence can be defined
with respect to other probability spaces (as well as for non-Boolean functions),
but we focus on the above definition.} in $\bitset^n$ and
 $x^{(\oplus i)}$ denotes $x$ with the $i^\xth$ bit flipped.
The {\it total influence\/} of a Boolean function $f$ (which we
simply refer to as {\em the influence\/} of $f$) is $I[f] = \sum_i{I_i[f]}$.

The study of the influence of a function and its individual influences
(distribution) has been the focus of many papers
(~\cite{BL,KKL,BKKKL,FK,Tal1,BK,Tal2,Fri1,Bop,Fri4,OSSS,DFKO} to mention a few --
for a survey see~\cite{KS}).
The influence of functions has played a central role in
several areas of computer science. In particular, this is true
for distributed computing (e.g.,~\cite{BL,KKL}),
hardness of approximation (e.g.,~\cite{DS,Khot}),
learning theory~(e.g.,~\cite{HM,BT,OS,OS2,DHKMRST})\footnote{Here we
referenced several works in which the influence appears explicitly.
The influence of variables plays an implicit role in many learning
algorithms, and in particular those that build on Fourier analysis,
beginning with~\cite{LMN}.}
 and
property testing~ (e.g.,~\cite{FKRSS,Bl1,Bl2,MORS2,RT}).
The notion of influence also arises naturally
in the context of probability theory (e.g.,~\cite{Rus,Tal3,BKS}),
game theory (e.g.,~\cite{Leh}), reliability theory~(e.g., \cite{KSV}),
as well as theoretical economics and political science
(e.g.,~\cite{Arrow,Kal1,Kal2}).

Given that the influence is such a 
basic measure of functions and it plays an important
role in many areas, we believe it is of interest to
study the algorithmic question of approximating
the influence of a function as efficiently as possible, that is
by querying the function on as few inputs as possible.
Specifically, the need for an efficient approximation for a function's
influence might arise in 
the design of sublinear algorithms, and in particular
property testing algorithms.

As we show, one cannot improve on a standard sampling argument for
the problem of estimating
 the influence of a general Boolean function,
which requires $\Omega(\frac{n}{I[f]})$ queries to the
function, for any constant multiplicative estimation factor.\footnote{
If one wants an {\em additive\/} error of $\eps$, then
$\Omega((n/\eps)^2)$ queries are necessary (when the influence is large)
\cite{MSW}.}
This fact justifies the study of 
subclasses of Boolean functions,
among which the family of monotone functions is a very natural
and central one.
Indeed, we show that the special structure
of monotone functions implies a useful behavior of their
influence, and thus the computational
problem of approximating the influence of such functions becomes
significantly easier.

\subsection{Our results and techniques}

We present a randomized
algorithm that approximates the 
influence of a monotone Boolean
function to within any multiplicative factor of $(1 \pm \epsilon)$
in $O\left(\frac{\sqrt{n}\log n}{I[f]} \poly(1/\eps) \right) $ expected
query complexity.
We also prove an almost matching lower bound of
$\Omega\left(\frac{\sqrt{n}}{\log n \cdot I[f]}\right)$
on the query complexity of any constant-factor approximation algorithm
for this problem (which holds for $I[f] = \Omega(1)$).

As noted above, the influence of a function can be
approximated by sampling random edges
(i.e., pairs $(x,x^{(\oplus i)})$ that differ on a single coordinate) from
the $\{0,1\}^n$ lattice. A random edge has
probability $\frac{I[f]}{n}$ to be influential (i.e,
satisfy $f(x) \neq f(x^{(\oplus i)})$), so a standard sampling
argument implies that it suffices to
ask $O(\frac{n}{I[f]}\poly(1/\eps))$
queries in order to approximate this probability
to within $(1\pm \eps)$.\footnote{We also note that in the case of
monotone functions, the total influence equals twice the sum of the Fourier
coefficients that correspond to singleton sets $\{i\}$, $i \in \{1,\ldots,n\}$.
Therefore, it is possible to approximate the influence of a function by
approximating this sum, which equals
$\frac{1}{2^{n}} \cdot \sum_{i=1}^n \left(\sum_{x\in\bitset^n:x_i=1} f(x) -
\sum_{x\in\bitset^n:x_i=0} f(x)\right)$.
However,
the direct sampling approach for such an approximation again requires $\Omega(n/I[f])$ samples.
}

In order to achieve better query complexity, we would like to increase
the probability of hitting an influential edge in a
single trial. The algorithm we present captures this intuition,
by taking random walks down the $\{0,1\}^n$ lattice\footnote{That is,
starting from a randomly selected point in $\{0,1\}^n$, at each step,
if the current point is $x$, we uniformly select an index $i$ such that
$x_i=1$ and continue the walk to $x^{(\oplus i)}$.},
and then averaging the total number of influential edges encountered in
all walks over the number of walks taken.
The crucial observation on which the algorithm relies, is that a
monotone function can have at most one influential edge in a single path,
and thus it is sufficient to query only the start and end points of the
walk to determine whether any influential edge was traversed.

Before continuing the technical discussion concerning the algorithm
and its analysis,
we make the following more conceptual note.
Random walks have  numerous applications in
Computer Science as they are an important tool for mixing
and sampling almost uniformly. In our context, where the walk is
performed on the domain of an unknown function, it is used for a
different purpose. Namely, by querying only the two endpoints of
a random walk (starting from a uniformly sampled element) we (roughly)
simulate
the process of taking a much larger sample of elements.

The main issue that remains is determining the length of the walk,
which we denote by $w$.
Let $p_w(f)$ denote the probability that a walk of length $w$
 (down the lattice and
from a uniformly selected starting point) passes through
some influential edge.\footnote{For technical reasons we actually consider a
slightly different measure than $p_w(f)$, but we ignore this technicality
in the introduction.}
 We are interested in analyzing  how $p_w(f)$ increases as a function of $w$.
We show that for $w$ that is $O(\eps\sqrt{n/\log n})$,
the value of $p_w(f)$ increases almost linearly with $w$.
Namely, it
 is  $(1\pm \eps)\cdot \frac{w}{n}\cdot I[f]$.
Thus, by taking $w$ to be  $\Theta(\eps\sqrt{n/\log n})$
we get an improvement by a factor of roughly $\sqrt{n}$ on the basic
sampling algorithm. We note though that by taking $w$ to be larger
we cannot ensure in general the same behavior of $p_w(f)$ as a function
of $w$ and $I[f]$, since the behavior might vary significantly depending on $f$.

The way we prove the aforementioned dependence of $p_w(f)$ on $w$
is roughly as follows.  For any edge $e$ in the Boolean lattice, let $p_w(e)$
denote the probability that a walk of length $w$ (as defined above)
passes through $e$. By the observation made previously, that a monotone
function can have at most one influential edge in a given path,
$p_w(f)$ is the sum of $p_w(e)$, taken over all edges $e$ that
are influential with respect to $f$.
For our purposes it is important that
$p_w(e)$ be roughly the same for almost all edges.
Otherwise, different functions that have the same number of influential edges,
and hence the same influence $I[f]$, but whose
influential edges are distributed differently in the Boolean lattice,
would give different values for $p_w(f)$. 
We show that for $w=O(\eps\sqrt{n/\log n})$,
the value of $p_w(e)$ increases almost linearly with $w$
for all but a negligible fraction of the influential
edges (where `negligible' is  with respect to $I[f]$).
This implies that $p_w(f)$ grows roughly linearly in $w$
for $w=O(\eps\sqrt{n/\log n})$.

To demonstrate the benefit of taking walks of length
$O(\sqrt{n})$, let us consider the
classic example of the {\it Majority\/}  function
on $n$ variables. Here, all influential edges are concentrated in the
exact middle levels of the lattice (i.e, all of
them are of the form $(x,x^{(\oplus i)})$ where the Hamming weight of $x$
is $\lfloor n/2 \rfloor$ or $\lceil n/2 \rceil$).
The probability, $p_w(e)$, of a walk of length $w$ passing through an influential
edge $e$ is simply the probability of starting the walk
at distance at most $w$ above the threshold $n/2$. Thus, taking longer
walks allows us, so to speak, to start our walk from
a higher point in the lattice, and still hit an influential edge. Since
the probability of a uniformly chosen point to fall
in each one of the the first $\sqrt{n}$ levels above the middle is
roughly the same, the probability of hitting an influential
edge in that case indeed grows roughly linearly in the size of the walk.
Nevertheless, taking walks of length which significantly exceeds
$O(\sqrt{n})$ (say, even $\Omega(\sqrt{n\cdot \log(n)})$) would add
negligible contribution to that probability (as this contribution is equivalent
to the probability of a uniformly chosen point to deviate
$\Omega(\sqrt{n\cdot \log(n)})$ levels
from the middle level) and thus the linear dependence on the
length of the walk is no longer preserved.

%

\section{Preliminaries}\label{prel.sec}
In the introduction we defined the influence of a function as
the sum, over all its variables, of their individual influence.
An equivalent definition is that the influence of a function $f$
is the expected
number of sensitive coordinates for a random input $x \in \bitset^n$
(that is, those coordinates $i$ for which $f(x) \neq f(x^{(\oplus i)})$).

It will occasionally be convenient to view $f$ as a $2$-coloring
of the Boolean lattice. Under this
setting, any ``bi-chromatic'' edge, i.e, an edge $(x,x^{(\oplus i)})$
such that $f(x) \neq f(x^{(\oplus i)})$,
will be called an {\it influential edge\/}.
The number of influential edges of a Boolean function $f$ is
$2^{n-1}\cdot I[f]$.\footnote{To verify this, observe that when
partitioning the Boolean lattice
into two sets with respect to a coordinate $i$, we end up with $2^{n-1}$ vertices in each
set. The individual influence of variable $i$, $I_i[f]$, is the fraction of the
``bi-chromatic'' edges among all
edges crossing the cut. Since
$I[f] = \sum_{i=1}^{n}I_i[f]$ we get that the total number of
influential edges is $2^{n-1}\cdot I[f]$.}

We consider the standard partial order `$\prec$' over the ($n$-dimensional) Boolean lattice.
Namely, for $x = (x_1,...,x_n), y = (y_1,...,y_n) \in \bitset^n$,
we use the notation $x \prec y$ to mean that $x_i \leq y_i$
for every $1 \leq i \leq n$, and $x_i < y_i$ for some
$1 \leq i \leq n$.
A Boolean function $f: \bitset^n \to \bitset $ is said to be
{\it monotone\/} if $f(x) \leq f(y)$  for all $x \prec y$.
A well known isoperimetric inequality implies that any monotone
Boolean function satisfies $I[f] = O(\sqrt{n})$ (see~\cite{FK} for a proof).
This bound is tight for the notable {\em Majority\/} function.

In this paper we deal mainly with monotone Boolean functions that have at least
constant Influence (i.e, $I[f] \geq c$, for some
$c \geq 0$), since the computational problem we study arises more naturally
when the function has some significant sensitivity.
As shown in~\cite{KKL}, the influence of a function is lower bounded
by $4\cdot\Pr[f = 1]\cdot\Pr[f=0]$, and so our analysis holds in particular
for functions that are not too biased (relatively balanced).



\noindent {\bf Notations.}  We use the notation $f(n) = \tilde{O}(g(n))$ if
$f(n) = O(g(n)\polylog(g(n)))$.
Similarly, $f(n) = \tilde{\Omega}(g(n))$ if
$f(n) = \Omega(g(n)/\polylog(g(n)) )$.

\section{The Algorithm}\label{alg.sec}

As noted in the introduction, we can easily get a
$(1\pm \eps)$-factor estimate of the influence with high constant probability
by uniformly sampling
$\Theta\left(\frac{n}{I[f]}\cdot \eps^{-2}\right)$
pairs $(x,x^{(\oplus i)})$ (edges in the Boolean lattice), querying the
function on these pairs, and considering the
fraction of influential edges observed in the sample.
We refer to this as the {\em direct sampling approach\/}.
However, since we are interested in an algorithm whose
complexity is  $\frac{\sqrt{n}}{I[f]}\cdot\poly(1/\eps)$
we take a different approach. To be precise, the algorithm
we describe works for $\eps$ that is above a certain
threshold (of the order of $\sqrt{\log n/n}$). However, if $\eps$ is
smaller, then
$\frac{n}{I[f]}\cdot \eps^{-2}$ is upper bounded by
$\frac{\sqrt{n}}{I[f]}\cdot\poly(1/\eps)$, and we can
take the direct sampling approach.
Thus we assume from this point on that
$\eps = \omega(\sqrt{\log n/n})$.


As discussed in the introduction,
instead of considering neighboring pairs, $(x,x^{(\oplus i)})$,
we consider pairs $(v,u)$ such that $v \succ u$
and there is a path down the lattice of length
roughly $\eps\sqrt{n}$ between $v$ and $u$. Observe that since the
function $f$ is monotone, if the path (down the lattice) from $v$ to $u$
contains an influential edge, then $f(v) \neq f(u)$, and
furthermore, any such path can contain at most one influential edge.
The intuition is that since we ``can't afford'' to detect influential
edges directly, we raise our probability of detecting edges by considering
longer paths.

In our analysis we show that this intuition can be formalized so as
to establish the correctness of the algorithm.
We stress that when considering a path, the algorithm only
queries its endpoints, so that it ``doesn't pay'' for the length of
the path.  The precise details  of the algorithm
are given in Figure~\ref{inf-alg.fig}.
When we say that we take a walk of a certain length $w$ down
the Boolean lattice {\em with a cut-off\/} at a certain
level $\ell$, we mean that we stop the walk (before taking
all $w$ steps) if we reach
a point in level $\ell$ (i.e., with Hamming weight $\ell$).

Note that $m$,
the number of walks taken, is a random variable. Namely, the algorithm
continues taking new walks until the number of ``successful'' walks (that is,
walks that pass through an influential edge) reaches a certain threshold,
which is denoted by $t$.
The reason for doing this, rather than deterministically setting the number of
walks and considering the random variable which is the number of successful walks,
is that the latter approach requires to know a lower bound on the influence
of $f$. While it is possible to search for such a lower bound (by working
iteratively in phases and decreasing the lower bound on the influence between phases)
our approach yields a somewhat simpler algorithm.


\begin{figure*}[htb]
\centerline{\fbox{\begin{minipage}{6.2in}
\BA{\hspace{-.5em}\rm: \bf  Approximating the Influence (given $\eps,\delta$ and
oracle access to $f$)}
\label{inf.alg}
\BE
\item
Set $\tilde{\eps} = \eps/4$,
$w = \frac{\tilde{\eps} \sqrt{n}}{16\sqrt{2\log(\frac{2n}{\tilde{\eps}})}}$,
$s^* = \frac{1}{2}\sqrt{2n\log(\frac{2n}{\tilde{\eps}})}$,
and $t = \frac{96\ln{(\frac{2}{\delta})}}{\eps^2}$.
   \item Initialize 
      $\alpha \leftarrow 0$,
      $m \leftarrow 0$,
     and $\hat{I} \leftarrow 0$.
   \item Repeat the following until $\alpha = t$:
   \BE
      \item Perform a random walk of length $w$ down the
         $\bitset^n$
         lattice from a uniformly chosen point
        $v$ with a cut-off
        at $n/2 - s^*-1$,
          and let $u$
      denote the endpoint of the walk.
        \item If $f(u) \neq f(v)$
            then $\alpha \longleftarrow \alpha+ 1$.
        \item $m \leftarrow m+ 1$
     \EE
    \item $\hat{I} \leftarrow \frac{n}{w}\cdot \frac{t}{m}$
\item Return $\hat{I}$.
\EE
\EA
\end{minipage}}}
\caption{\small The algorithm for approximating the influence of a function $f$.}
\label{inf-alg.fig}
\end{figure*}

In what follows we assume for simplicity that $I[f] \geq 1$. As we discuss subsequently,
this assumption can be easily replaced by $I[f] \geq c$ for any constant $c >0$,
or even $I[f] \geq n^{-c}$, by performing a slight modification in the setting
of the parameters of the algorithm. 
\BT\label{inf-alg.thm}
\sloppy
For every monotone function $f : \bitset^n \to \bitset$ such that $I[f] \geq 1$,
and for every  $\delta > 0$
and $\eps = \omega(\sqrt{\log n/n})$,
with probability at least $1-\delta$, the output, $\hat{I}$,
of Algorithm~\ref{inf.alg} satisfies:
$$(1-\eps)\cdot I[f] \leq \hat{I} \leq (1+\eps)\cdot I[f]\;.$$
Furthermore,
with probability at least $1-\delta$, the number
of queries performed by the algorithm  is
$O\left(\frac{\log(1/\delta)}{\eps^3}\cdot \frac{\sqrt{n}\log(n/\eps)}{I[f]}\right)$.
\ET
We note that the (probabilistic) bound on the number of queries performed by the
algorithm implies that the expected query complexity of the algorithm
is $O\left(\frac{\log(1/\delta)}{\eps^3}\cdot \frac{\sqrt{n}\log(n/\eps)}{I[f]}\right)$.
Furthermore, the probability that the algorithm performs a number of
queries that is more than $k$ times the expected value decreases exponentially
with $k$.

\smallskip
The next definition is central to our analysis.
\BD\label{p-w-s.def}
For a (monotone) Boolean function $f$ and integers $w$ and $s^*$, let
$p_{w,s^*}(f)$ denote the probability that a random walk of length $w$
down the Boolean lattice, from a uniformly selected point
and with a cut-off at $n/2-s^*-1$,
starts from $f(v) = 1$ and reaches
$f(u) = 0$.
\ED

Given the definition of $p_{w,s^*}(f)$, we next state and prove the main lemma
on which the proof of Theorem~\ref{inf-alg.thm} is based.
\BL\label{inf-alg.lem}
Let $f$ satisfy $I[f] \geq 1$,
let $\eps > 0$ satisfy
$\eps > \frac{8\sqrt{2\log(\frac{8n}{\eps})}}{\sqrt{n}}$,
and denote $\tilde{\eps} = \eps/4$.
For any
$w \leq \frac{\tilde{\eps} \sqrt{n}}{16\sqrt{2\log( \frac{2n}{\tilde{\eps} I[f]})}}$
and for
$s^* = \frac{1}{2}\sqrt{n}\cdot\sqrt{2\log(\frac{2n}{\tilde{\eps}})}$
we have that
$$(1-\eps/2)\cdot \frac{w}{n}\cdot I[f]
   \;\leq\; p_{w,s^*}(f) \;\leq\; (1+\eps/2)\cdot \frac{w}{n}\cdot I[f]\;. $$
\EL

\BPF
For a point $y \in \bitset^n$, let $h(y)$ denote its Hamming weight (which we
also refer to as the {\em level\/} in the Boolean lattice that it belongs to).
By the choice of
$s^* = \frac{1}{2}\sqrt{n}\sqrt{2\log(\frac{2n}{\tilde{\eps}})}$,
and since $I[f] \geq 1$,
the number of points $y$  for which
$h(y) \geq n/2 + s^*$ or $h(y) \leq n/2 - s^*$, is upper bounded by
$2^n \cdot \frac{\tilde{\eps} I[f]}{n}$.
\ifnum\conf=0
 Each such point $y$ is incident to $n$ edges, and each edge has
two endpoints.
\fi
It follows that there are at most
$2^{n-1}\cdot \tilde{\eps}I[f]$ edges $(y,x)$ for which
$h(y) \geq n/2 + s^*$ or $h(y) \leq n/2 - s^*$.
Recall that an influential edge $(y,x)$ for $h(y) = h(x)+1$,  is an edge that satisfies
$f(y) = 1$ and $f(x)=0$.
Let $e_{s^*}(f)$ denote the number of influential edges $(y,x)$
such that $n/2 - s^* \leq h(x),h(y) \leq n/2+s^*$.
Since the total number of influential edges is  $2^{n-1} I[f]$,
we have that
\begin{equation} \label{e-s-def.eq}
(1-\tilde{\eps})\cdot 2^{n-1} I[f]
\;\leq\; e_{s^*}(f) \;\leq \;
2^{n-1} I[f] \;.
\end{equation}

Consider any influential edge $(y,x)$ where
$h(y) = \ell$ and $\ell \geq n/2 - s^*$.
We are interested in obtaining bounds on the probability that
a random walk of length $w$
(where
$w \leq \frac{\tilde{\eps} \sqrt{n}}{16\sqrt{2\log( \frac{2n}{\tilde{\eps} I[f]})}}$)
down the lattice, starting from a uniformly selected point $v \in \bitset^n$,
and with a cut-off at $n/2 - s^*-1$,
passes through $(y,x)$.
First, there is the event that $v = y$
and the edge $(y,x)$ was selected in the first step of the walk.
This event occurs with probability
$2^{-n}\cdot \frac{1}{\ell}$. Next there is the event that $v$ is at
distance $1$ from $y$ (and above it, that is, $h(v) = h(y)+1 = \ell+1$),
and the edges $(v,y)$ and $(y,x)$
are selected. This occurs with probability
$2^{-n} \cdot (n-\ell) \cdot \frac{1}{\ell+1}\cdot \frac{1}{\ell}$.
In general, for every $1 \leq i \leq w-1$ we have $(n-\ell)\cdots (n-\ell-i+1)$
pairs $(v,P)$ where $v\succ y$ and $w(v) = \ell+i$, and where $P$ is
a path down the lattice from $v$ to $y$. The probability of selecting
$v$ as the starting vertex is $2^{-n}$ and the probability of taking the
path $P$ from $v$ is $\frac{1}{(\ell+i)\cdots(\ell+1)}$.
Therefore, the probability that the random
walk passes through $(y,x)$ is:
\begin{equation}\label{prob-pass.eq}
2^{-n}\cdot \frac{1}{\ell} \cdot
\left(1 + \sum_{i=1}^{w-1} \frac{(n-\ell)\cdots (n-\ell-i+1)}
                                {(\ell+i)\cdots(\ell+1)}     \right)
 = 2^{-n}\cdot \frac{1}{\ell} \left(1 + \sum_{i=1}^{w-1}
             \prod_{j=0}^{i-1}\frac{n-\ell-j}{\ell+i-j}\right) \;.
\end{equation}
Let $\ell = n/2 + s$ (where $s$ may be negative), and denote
$\tau(\ell,i,j) \eqdef \frac{n-\ell -j}{\ell+i - j}$. Then
\BEQ\label{star-def.eq}
\tau(\ell,i,j)
  = \frac{n/2 -s -j}{n/2 +s + i - j}
  = 1 - \frac{2s + i}{n/2 + s + i - j} \;.
\EEQ
Consider first the case that $\ell \geq n/2$, i.e $\ell = n/2 + s$ ($s \geq 0$).
In that case it is clear that $\tau(\ell,i,j)  \leq 1$ (since $j \leq i$),
so $\prod_{j=0}^{i-1}\tau(\ell,i,j)$ is upper bounded by 1.
In order to lower bound $\prod_{j=0}^{i-1}\tau(\ell,i,j)$,
we note that
\BEQ
\tau(\ell,i,j) \geq 1 - \frac{2s+w}{n/2} = 1- \frac{2(2s+w)}{n}\;.
\EEQ
Thus, for $s \leq s^*$ we have
\begin{eqnarray}
\prod_{j=0}^{i-1}\tau(\ell,i,j)
&\geq& \prod_{j=0}^{i-1}\left(1- \frac{2(2s+w)}{n}\right) \nonumber \\
&\geq& \left(1- \frac{2(2s+w)}{n}\right)^w    \hspace{3.0 pc}
                \mbox{(since $i \leq w$)} \nonumber \\
&\geq& 1- \frac{2(2s+w)w}{n}
                     \nonumber \\
&\geq& 1- \frac{6s^*w}{n}          \hspace{6.0 pc}
            \mbox{($2s+w \geq 3s^*$ since $s \leq s^*$ and $w \leq s^*$)} \nonumber \\
&=& 1- \frac{3\tilde{\eps}}{16}    \hspace{7.0 pc}
          \mbox{(by the definitions of $s^*$ and $w$)}
\nonumber \\
&\geq& 1- \tilde{\eps}/2 \;.
\end{eqnarray}
%
%
Therefore, we have that for $n/2 \leq \ell \leq n/2+s^*$,
\BEQ \label{ell-above.eq}
1- \tilde{\eps}/2 \;\leq\; \prod_{j=0}^{i-1}\frac{n-\ell-j}{\ell+i-j} \;\leq\; 1 \;,
\EEQ
and for $\ell > n/2+s^*$ it holds that
\BEQ\label{ell-big.eq}
\prod_{j=0}^{i-1}\frac{n-\ell-j}{\ell+i-j} \leq 1\;.
\EEQ

\bigskip\noindent
We turn to the case where $n/2- s^* \leq \ell < n/2$.
Here we have
\BEQ
\tau(\ell,i,j) \;=\;
    1+ \frac{2s-i}{n/2-s+i-j} \; \geq\;  1- \frac{2w}{n - 2w}
   \;\geq\;  1- \frac{4w}{n}
\EEQ
 where the last inequality follows from the fact that
$w < n/4$.
Thus,
\BEQ
\prod_{j=0}^{i-1}\tau(\ell,i,j) \;\geq\; \left(1- \frac{4w}{n}\right)^w \;\geq\;
1- \frac{4w^2}{n} \;=\; 1 - \frac{4}{n}\cdot
\left(\frac{\tilde{\eps} \sqrt{n}}{16\sqrt{2\log(\frac{2n}{\tilde{\eps} I[f]})}}\right)^2
\;>\; 1- \tilde{\eps}^2/2 \;>\; 1- \tilde{\eps}/2\;.
\EEQ
On the other hand,
\BEQ
\tau(\ell,i,j) \;=\; 1+ \frac{2s-i}{n/2-s+i-j} \;\leq\;  1 + \frac{2s}{n/2 - s}
\;\leq\; 1 + \frac{8s^*}{n}\;,
\EEQ
where the last inequality 
holds since $n \geq 2s$.
Thus, we have
\BEQ
\prod_{j=0}^{i-1}\tau(\ell,i,j) \;\leq\; \left(1+ \frac{8s^*}{n}\right)^w \;\leq\;
1+ \frac{16s^*w}{n} \;=\; 1+ \tilde{\eps}/2 \;.
\EEQ
where the second 
inequality follows from the inequality
$(1+\alpha)^k \leq 1 + 2\alpha k$ which holds for $\alpha < 1/(2k)$;
Indeed, in our case $8s^*/n \leq 1/(2w)$ (this is equivalent to $w \leq n/16s^*$ which
holds given our setting of $s^*$ and the upper bound on $w$).

\noindent
We therefore have that for $n/2 - s^* \leq \ell < n/2$,
\BEQ \label{ell-below.eq}
1- \tilde{\eps}/2 \;\leq\; \prod_{j=0}^{i-1}\frac{n-\ell-j}{\ell+i-j}
\;\leq\; 1+ \tilde{\eps}/2  \;.
\EEQ
Combining Equations~(\ref{ell-above.eq}) and~(\ref{ell-below.eq}), we have that
for $n/2 - s^* \leq \ell \leq n/2+s^*$,
\BEQ \label{ell-above-below.eq}
1- \tilde{\eps}/2 \;\leq\; \prod_{j=0}^{i-1}\frac{n-\ell-j}{\ell+i-j}
   \;\leq\; 1+ \tilde{\eps}/2  \;.
\EEQ
Now, we are interested in summing up the probability, over all random walks,
that the walk passes through an influential edge.
Since the function is monotone, every random walk passes through at most one
influential edge, so the sets of random walks that correspond to
different influential edges are disjoint (that is, the event that
a walk passes through an influential edge $(y,x)$ is disjoint from the
event that it passes through another influential edge $(y',x')$).
Since the edges that contribute to $p_{w,s^*}(f)$ are all from levels
$\ell \geq n/2 - s^*$ (and since there are $2^{n-1}I[f]$ influential edges
in total), by Equations~(\ref{prob-pass.eq}),~(\ref{ell-big.eq})
and~(\ref{ell-above-below.eq})
we have
\begin{eqnarray}
p_{w,s^*}(f)
&\leq& 2^{n-1}I[f]2^{-n} \cdot \frac{1}{n/2 - s^*}
 \left(1+ \sum_{i=1}^{w-1}(1+ \tilde{\eps}/2)\right)  \label{pwf-ub1.eq}\\
&\leq& \frac{1}{2}I[f] \cdot \frac{1}{n/2 - s^*}\cdot w(1+ \tilde{\eps}/2)  \\
&\leq& \frac{1}{2}I[f] \cdot \frac{2}{n}(1+ \tilde{\eps})\cdot w(1+ \tilde{\eps}/2)
                    \label{eps-lb.eq} \\
&\leq& \frac{I[f]\cdot w}{n} \cdot (1+ 2\tilde{\eps})  \\
&=& \frac{I[f]\cdot w}{n}(1+ \eps/2)  \;,
\label{pwf-ub.eq}
\end{eqnarray}
where Equation~(\ref{eps-lb.eq}) follows from the definition of
$s^*$, the premise of the
lemma that $\eps > \frac{8\sqrt{2\log(\frac{8n}{\eps})}}{\sqrt{n}}$ and
$\tilde{\eps} = \eps/4$.

\medskip\noindent
For lower bounding $p_{w,s^*}(f)$, we will consider only the contribution of
the influential edges that belong to levels
$\ell \leq n/2 +s^*$. Consequently, Equations~(\ref{e-s-def.eq}),~(\ref{prob-pass.eq})
and~(\ref{ell-above-below.eq})  give in total
\begin{eqnarray} p_{w,s^*}(f) 
&\geq& 2^{n-1}(1- \tilde{\eps})I[f]2^{-n}
 \cdot \frac{1}{n/2 + s^*}\left(1+ \sum_{i=1}^{w-1}(1- \tilde{\eps}/2)\right) \\
&\geq& \frac{1}{2}I[f](1-\tilde{\eps})w(1-\tilde{\eps/2}) \cdot \frac{1}{n/2 + s^*} \\
&\geq& \frac{1}{2}I[f]\cdot w(1-\tilde{\eps})(1-\tilde{\eps/2})
            \cdot \frac{2}{n}(1- \tilde{\eps}) \label{eps-lb2.eq} \\
&\geq& \frac{I[f]\cdot w}{n}(1-2\tilde{\eps}) \label{eps-manip.eq} \\
&=& \frac{I[f]\cdot w}{n}(1- \eps/2)\;,
\label{pwf-lb.eq}
 \end{eqnarray}
where Equation~(\ref{eps-lb2.eq}) follows from the definition of
$s^*$, the premise of the
lemma that $\eps > \frac{8\sqrt{2\log(\frac{8n}{\eps})}}{\sqrt{n}}$ and
$\tilde{\eps} = \eps/4$.

\medskip\noindent
Equations~(\ref{pwf-ub.eq}) and~(\ref{pwf-lb.eq}) give
\BEQ\label{p-w-s-bounds.eq}
(1-\eps/2)\cdot \frac{w}{n}\cdot I[f] \leq p_{w,s^*}(f) \leq
    (1+\eps/2)\cdot \frac{w}{n}\cdot I[f]\;,
\EEQ
 as claimed in the Lemma.
\EPF

\BPFOF{Theorem~\ref{inf-alg.thm}}
For $w$ and $s^*$ as set by the algorithm, let
$p_{w,s^*}(f)$ be as in Definition~\ref{p-w-s.def},
where we shall use the shorthand $p(f)$.
Recall that $m$ is a random variable denoting the number of iterations
performed by the algorithm until it stops (once $\alpha = t$).
Let
$\tilde{m} = \frac{t}{p(f)}$, $\tilde{m}_1 = \frac{\tilde{m}}{(1 + \eps/4)}$,
and  $\tilde{m}_2= \frac{\tilde{m}}{(1 - \eps/4)}$.
We say that an iteration of the algorithm is {\em successful\/}
if the walk taken in that iteration passes through an influential edge
(so that the value of $\alpha$ is increased by $1$).
Let $\hat{p}(f) = \frac{t}{m}$ denote the fraction of successful iterations.

Suppose that $\tilde{m}_1 \leq m \leq \tilde{m}_2$.
In such a case,
\BEQ\label{est-frac.eq}
(1- \eps/4)\cdot p(f) \;\leq\; \hat{p}(f) \;\leq\; (1+ \eps/4)p(f)
\EEQ
since $\hat{p}(f) = \frac{t}{m} = \frac{p(f)\cdot \tilde{m}}{m}$.
By the definition of the algorithm,
 $\hat{I} = \frac{n}{w}\cdot \frac{t}{M} = \frac{n}{w}\cdot \hat{p}(f)$ so by
Lemma~\ref{inf-alg.lem} (recall that by the
premise of the theorem, $\eps = \omega(\sqrt{\log n/n})$)
we have
\BEQ\label{estimate-I.eq}
(1- \eps)I[f]
\;\leq\; (1- \eps/2)(1- \eps/4)I[f] \;\leq\;
 \hat{I} \;\leq\; (1+ \eps/4)(1+ \eps/2)I[f] \;\leq\; (1+ \eps)I[f]
\EEQ
and thus (assuming $\tilde{m}_1 \leq m \leq \tilde{m}_2$),
 the output of the algorithm provides the estimation we are looking for.

\medskip
 It remains to prove that
 $\tilde{m}_1 \leq m \leq \tilde{m}_2$ with probability
at least $1-\delta$.
Let $X_i$ denote the indicator random variable whose value is $1$ if and only
if the $i^\xth$ iteration of the algorithm was successful, and
let $X = \sum_{i=1}^{\tilde{m}_1} X_i$. By the definition of $X_i$,  we have that
$\E[X_i] = p(f)$, and so
(by the definition of $\tilde{m}_1$ and $\tilde{m}$) we have that
$\E[X] = \tilde{m}_1\cdot p(f) = \frac{t}{1+ \eps/4}$
Hence, by applying the multiplicative Chernoff bound,
\BEQ\label{lower-dev.eq}
\Pr[m < \tilde{m}_1]
   = \Pr[X > t] = \Pr[X > (1+ \eps/4)\E[X]]
   \leq \exp\left(-\frac{1}{3}\left(\frac{\eps}{4}\right)^2\frac{t}{1+ \eps/4}\right)
\leq \exp\left(-\frac{\eps^2t}{96}\right)
\EEQ
Thus, for $t = \frac{96\ln{(\frac{2}{\delta})}}{\eps^2}$ we have that
$\Pr[m < \tilde{m}_1] \leq \frac{\delta}{2}$.
By an analogous argument we get that
$\Pr[m > \tilde{m}_2] \leq \frac{\delta}{2}$,
and so $\tilde{m}_1 \leq m \leq \tilde{m}_2$ with probability
at least $1-\delta$, as desired.
%
%
%
%
%
%
%

Since we have shown that $m \leq \tilde{m}_2$ with probability at least $1-\delta$,
and the query complexity of the algorithm is $O(m)$, we have that, with
probability at least $1-\delta$, the query complexity
is upper bounded by
%
%
\BEQ
O(\tilde{m}_2)  \;=\; O\left(\frac{t}{p(f)}\right)  \;=\;
  O\left(\frac{t\cdot n}{w\cdot I[f]}\right)
  \;=\; O\left(\frac{\log(1/\delta)}{\eps^3}\cdot\frac{\sqrt{n}\log(n/\eps)}{I[f]}\right)\;,
\EEQ
as required.
\EPFOF

\paragraph{Remark.}
We assumed that $I[f] \geq 1$  only for the sake of technical simplicity.
This assumption can be replaced with $I[f] \geq \frac{1}{n^c}$ for any constant
$c \geq 0$, and the only modifications needed in the algorithm and its analysis
are the following.
The level of the cutoff $s^*$  should be set to
$s^* = \sqrt{n/2}\cdot \sqrt{\log(\frac{2n}{\tilde{\eps}n^{-c}})}
  = \frac{1}{2}\sqrt{n}\sqrt{2c\log(2n)+\log(1/\tilde{\eps})} $
(which is a constant factor larger than the current setting),
and the length $w$ of the walks in the algorithm  should be set to
$w = \frac{\tilde{\eps} \sqrt{n}}{16\sqrt{2\log( \frac{2n}{\tilde{\eps} n^{-c}})}} $
(which is a constant factor smaller than the current setting).

The first modification follows from the fact that the number of points 
$y$ whose Hamming weight $h(y)$ is at least $n/2 + r\cdot \sqrt{n/2}$ or at most 
$n/2 - r\cdot \sqrt{n/2}$ is upper bounded by $2^n \cdot 2e^{-r^2}$.
This implies that the number of edges $(y,x)$ (where $h(y) = h(x)+1$) such that  
$h(y) \geq n/2 + r\cdot \sqrt{n/2}$ or $h(y) \leq n/2 - r\cdot \sqrt{n/2}$ is
upper bounded by $n\cdot 2^{n} \cdot 2^{-r^2}$.
Requiring that
the latter is no more than $\tilde{\eps}\cdot I[f]2^{n-1} \geq \tilde{\eps}\cdot n^{-c} 2^{n-1} $ 
(i.e, $\tilde{\eps}$-fraction of the total number of influential edges), yields the desired 
$r$, where $s^* = r\sqrt{n/2}$. 
The second modification, i.e, in the length of the walk, 
is governed by the choice of $s^*$,
since, by the analysis, their product should be bounded by $O(\tilde{\eps} n)$. 
Since in both expressions
$1/I[f] = n^c$ appears only inside a $\log$ term, this translates only to constant factor 
increase.

We note that the lower bound we give in Section~\ref{lb.sec} applies
only to functions with (at least) constant influence, and so in the above case where
$I[f] = 1/\poly(n)$, the tightness of the algorithm (in terms of query complexity) is
not guaranteed.
\section{A Lower Bound}\label{lb.sec}

\newcommand{\maj}{{\rm maj}}

In this section we prove a lower bound of
$\Omega\left(\frac{\sqrt{n}}{I[f]\cdot \log n}\right)$
on the query complexity of approximating the influence
of monotone functions. Following it we explain how a related construction
gives a lower bound of $\Omega\left(\frac{n}{I[f]}\right)$
on approximating the influence of {\em general\/} functions.
The idea for the first lower bound is the following. We show that
any algorithm that performs $o\left(\frac{\sqrt{n}}{I[f]\cdot \log n}\right)$
queries cannot
distinguish with  constant success probability
between that following: (1) A certain threshold function
(over a relatively small number of variables), and (2) A function selected
uniformly at random from a certain family of functions that have
significantly higher influence than the threshold function. The functions
in this family can be viewed as ``hiding their influence behind
the threshold function''.  More precise details follow.

We first introduce
one more notation. For any integer $1 \leq k \leq n$ and $0 \leq t \leq k$, let
$\tau^t_k : \bitset^n \to \bitset$ be the
{\em $t$-threshold function\/} over $x_1,\ldots,x_k$.
That is, $\tau^t_k(x) =1$ if and only if $\sum_{i=1}^k x_i \geq t$.
 Observe that  (since for every $1 \leq i\leq k$ we have that
 $I_i[\tau^t_k] = 2^{-k}\cdot 2\cdot {k-1 \choose t-1 }$ while for
 $i > k$ we have that $I_i[\tau^t_k] = 0$),
$I[\tau^t_k] = k \cdot 2^{-(k-1)} \cdot  {k-1 \choose t-1}$.

The above observation implies that
 for every sufficiently large $k$
($k \geq 2\log n$ suffices),  there exists a setting of
$t < k/2$, which
we denote by $t(k,1)$, such that $I[\tau^{t(k,1)}_k] = 1-o(1)$
(where the $o(1)$ is with respect to $k$).
This setting satisfies ${k-1 \choose t(k,1)-1} = \Theta(2^k/k)$
(so that $t(k,1) = k/2 - \Theta(\sqrt{k\log k}$)).

\BT\label{inf-lb.thm}
For every $I^*$ such that
$2 \leq I^* \leq \sqrt{n}/\log n$,
there exists a family of monotone functions $F_{I*}$
such that $I[f] \geq I^*$ for every
$f\in F_{I^*}$, but any algorithm
that distinguishes with probability at least $2/3$
between a uniformly selected function in $F_{I^*}$ and
$\tau^{t(k,1)}_k$ for $k = 2\log n$,
 must perform $\Omega\left(\frac{\sqrt{n}}{I^*\cdot \log n}\right)$
queries.
\ET
In particular, considering $I^* = c$ for any constant $c \geq 2$,
we get that every algorithm for approximating the influence to
within a multiplicative factor of $\sqrt{c}$
must perform $\tilde{\Omega}(\sqrt{n})$ queries.
If we increase the lower bound on the influence,
then the lower bound on the complexity of the algorithm decreases,
but the approximation factor (for which the lower bound holds),
increases. We note that the functions for which the lower bound
construction hold are not balanced, but we can easily make them
very close to balanced without any substantial change in the
argument (by ``ORing'' $\tau^{t(k,1)}_k$ as well as every function
in $F_{I^*}$ with $x_1$).
We also note that for $I^* = \Omega(\sqrt{\log n})$
we can slightly improve the lower bound on approximating
the influence to
   $\Omega\left(\frac{\sqrt{n}}{I^*\cdot\sqrt{\log(\sqrt{n}/I^*)}}\right)$
(for a slightly smaller approximation factor).
We address this issue following the proof.

\smallskip
\BPF
For $k = 2\log n$ and for any $0 \leq t \leq k$, let
$L^t_k\eqdef \{x\in \bitset^k\;:\; \sum_{i=1}^k x_i = t\}$.
We shall also use the shorthand $\tilde{t}$ for $t(k,1)$.
Fixing a choice of $I^*$, each function in $F_{I^*}$ is
defined by a subset $R$ of $L^{\tilde{t}}_k$
where $|R| = \beta(I^*)\cdot 2^k$
for $\beta(I^*)$ that is set subsequently.
We denote the corresponding function by $f_R$ and define it as follows:
For every $x \in \bitset^n$, if
$x_1\ldots x_k \notin  R$,
then $f_R(x) = \tau^{\tilde{t}}_k(x)$, and
if  $x_1\ldots x_k \in R$, then
$f_R(x) = \maj'_{n-k}(x)$, where
$\maj'_{n-k}(x) =1$ if and only if  $\sum_{i=k+1}^n x_i > (n-k)/2$.
By this definition, for every $f_R \in F_{I^*}$
\BEQ
I[f_R]  \geq 
       \beta(I^*)\cdot I[\maj'_{n-k}] \;.
\EEQ
If we take $\beta(I^*)$ to be
$ \beta(I^*)  =  I^*/I[\maj'_{n-k}] = c I^* /\sqrt{n-k}$
(for $c$ that is roughly $\sqrt{\pi/2}$),
then in $F_{I^*}$ every function has influence  at least $I^*$.
Since $\beta(I^*)$ is upper bounded by $|L^{\tilde{t}}_k|/2^k$, which is
of the order of $k/2^k = 2\log n/2^k$, this construction is
applicable to $I^* =  O(\sqrt{n}/\log n)$.

Consider an algorithm that needs to distinguish between $\tau^{\tilde{t}}_k$
and a uniformly selected $f_R \in F_{I^*}$.
Clearly, as long as the algorithm doesn't perform a query on
$x$ such that $x_1\ldots x_k \in R$,
the value returned by $f_R$ is the same as that of $\tau^{\tilde{t}}_k$.
But since $R$ is selected uniformly in $L^{\tilde{t}}_k$, as long as
the algorithm performs less than
$\frac{|L^{\tilde{t}}_k|}{c'\cdot \beta(I^*)\cdot 2^k}$ queries
(where $c'$ is some sufficiently
large constant), with high constant probability (over the choice of $R$),
it won't ``hit'' a point in $R$.
Since
$\frac{|L^{\tilde{t}}_k|}{c'\cdot \beta(I^*)\cdot 2^k} =
       \Theta\left(\frac{\sqrt{n}}{\log n \cdot I^*}\right)$,
the theorem follows.
\EPF

\medskip
In order to get the  aforementioned slightly higher lower bound for 
$I^* = \Omega(\sqrt{\log n})$,
we modify the settings in the proof of Theorem~\ref{inf-lb.thm}
in the following manner. We set $k = \log(\sqrt{n}/I^*)$ and $t = k/2$
(so that the ``low influence'' function is simply a majority function 
over $k$ variables, $\tau_k^{k/2}$).
For the ``high influence'' function, we let $R$ consist of
a single point $\tilde{x}$ in $L_k^{k/2}$, where for each $R = \{\tilde{x}\}$
we have a different function in $F_{I^*}$ (as defined in the proof of
Theorem~\ref{inf-lb.thm}). It follows that for each such $R$,
$I[f_R] = (1-o(1))\sqrt{k} + \frac{1}{2^k}\sqrt{n-k} \geq I^*$,
while  $I[\tau_k^{k/2}] \approx \sqrt{k} = O(\sqrt{\log{n}})$.
By the same argument as in the proof of Theorem~\ref{inf-lb.thm},
if the algorithm preforms less than 
$\frac{c'|L_k^{k/2}|}{|R|} = \frac{2^k}{{c'\sqrt{k}}} =
\frac{\sqrt{n}}{c'I^*\sqrt{\log({\frac{\sqrt{n}}{I^*}})}}$ queries
(for small enough $c'$), with high probability it won't ``hit'' $\tilde{x}$, 
and thus will not be able to distinguish
between a randomly selected function $f \in f_R$ (where the randomness is over the choice of
$\tilde{x} \in L_k^{k/2}$) and $\tau_k^{k/2}$. 

\paragraph{A lower bound of $\Omega(n/I[f])$ for general functions.}
We note that for general (not necessarily monotone) functions, there
is a lower bound of $\Omega(n/I[f])$ on estimating the influence, which
implies that it is not possible in general to improve on the simple
edge-sampling approach (in terms of the dependence on $n$ and $I[f]$).
Similarly to what we showed in the case of monotone functions, we show
that for every $I^* \geq 2$, it is hard to distinguish
between the dictatorship function $f(x) = x_1$ (for which $I[f]=1$)
and a uniformly selected function in a family $F_{I^*}$ of functions,
where every function in $F_{I*}$ has influence at least $I^*$.

Similarly to the construction in the proof of Theorem~\ref{inf-lb.thm},
we consider the first $k$ variables, where here $k = \log n$.
Fixing $I^*$ (where $I^* = o(n)$ or else the lower bound is trivial),
each function in $F_{I^*}$ is defined by a subset $R$ of $\bitset^k$
such that $|R| = I^*$.
We denote the corresponding function by $f_R$ and define it as follows:
For every $x \in \bitset^n$, if
$x_1\ldots x_k \notin R$ then $f_R(x) = x_1$, and
if $x_1\ldots x_k \in R$, we let $f_R(x) = \bigoplus_{i=k+1}^n x_i$.
By this definition (since $2^k = n$), for every $f_R \in F_{I^*}$
$I[f_R]  \geq (1-2I^*/n) + (I^*/n)\cdot (n-k) \geq I^*$.
The argument for establishing that it is hard to distinguish
between $f(x) = x_i$ and a uniformly selected function in
$F_{I^*}$ is essentially the same as in the proof of Theorem~\ref{inf-lb.thm}.



\newpage

\bibliographystyle{plain}
\bibliography{inf}

\end{document}